\documentclass{PoS}

\usepackage{amsfonts}
\usepackage{amsmath}
\usepackage{subfigure} 

\newcommand{\TB}{\textrm{B}}

\newcommand{\TP}{\textrm{P}}
\newcommand{\TV}{\textrm{V}}
\newcommand{\TA}{\textrm{A}}
\newcommand{\TT}{\textrm{T}}

\title{
  Determination of $B_K$
  using improved staggered fermions (I):
  SU(3) chiral perturbation theory fit
}

\ShortTitle{
  Determination of $B_K$ using improved staggered fermions: (I) 
SU(3) fit
}

\author{\speaker{Taegil Bae}, Hyung-Jin Kim, Jangho Kim,
  Jongjeong Kim, Kwangwoo Kim, Boram Yoon, Weonjong Lee\\
  Frontier Physics Research Division and Center for Theoretical Physics \\
  Department of Physics and Astronomy, 
  Seoul National University, Seoul, 151-747, South Korea \\
  E-mail: \email{wlee@snu.ac.kr}}

\author{Chulwoo Jung \\
  Physics Department, Brookhaven National Laboratory,
  Upton, NY11973, USA \\
  E-mail: \email{chulwoo@bnl.gov}}

\author{Stephen R. Sharpe\\
  Physics Department, University of Washington, Seattle, WA 98195-1560 \\
  E-mail: \email{sharpe@phys.washington.edu}}

\abstract{ We present the results for $B_K$ calculated using HYP-smeared
staggered fermions using lattices generated by the MILC collaboration using the
asqtad staggered action.
We have done the calculation on 8 ensembles of these lattices,
including three different lattice spacings ($a=0.12, 0.09, 0.06$ fm). 
We fit the data to forms based on those predicted by
SU(3) mixed-action partially-quenched staggered chiral perturbation theory.
Our preliminary result is 
$ B_K(\text{NDR},\mu=2\text{ GeV}) = 0.528   \pm 0.011 \pm 0.048 $,
where the first error is statistical and the second systematic.
The error turns out to be larger than that from an analysis
using SU(2) chiral perturbation theory.
}

\FullConference{
  The XXVII International Symposium on Lattice Field Theory - LAT2009\\
  July 26-31 2009
  Peking University, Beijing, China
}

\begin{document}

\section{Introduction} 
Indirect CP violation violation in the neutral kaon system
and is conventionally parameterized by 
$\varepsilon = (2.228 \pm 0.011) \times 10^{-3}$.
In the Standard Model, this experimental parameter is related to the
matrix element of a $\Delta S=2$ four-fermion operator, itself
parameterized by 
\begin{equation}
B_K = \frac{\langle K_0 | \bar s \gamma_\mu^L d \bar s \gamma_\mu^L d | K_0\rangle}
{(8/3) f_K^2 M_K^2}\,.
\end{equation}
The relation between experiment and theory takes the form
(see, e.g., Ref.~\cite{buras-1998,buras-2008})
\begin{eqnarray}
\varepsilon &=& \exp(i\phi_\varepsilon) \
      \sin(\phi_\varepsilon)\left[ \sqrt{2} C_\varepsilon \
                  {\rm Im}\lambda_t \ X \ \hat{B}_K  + \xi \right]
+\textrm{dim-8 contributions} \,,
\label{eq:epsK}
\end{eqnarray}
where $\phi_\varepsilon$ is the (well-determined) phase of $\epsilon_K$,
\begin{equation}
      C_\varepsilon = \frac{G_F^2 F_K^2 m_K M_W^2}
      {6 \sqrt{2} \pi^2 \Delta M_K} \,,\qquad 
      \lambda_i = V_{is}^* V_{id}\,, 
\end{equation}
and the dependence on charm and top masses enters through the Inami-Lin
functions $S_{0,3}$
\begin{eqnarray}
      X &=& {\rm Re} \lambda_c [ \eta_1 S_0(x_c) - \eta_3 S_3(x_c,x_t) ]
      - {\rm Re} \lambda_t \eta_2 S_0(x_t) \,,\quad 
      x_i = m_i^2 / M_W^2 \,, 
\end{eqnarray}
with $\eta_{1-3}$ Wilson coefficients.
The $\xi = {{\rm Im}A_0}/{{\rm Re}A_0}$ term in (\ref{eq:epsK})
provides a small correction
to the proportionality of $\epsilon_K$ and $B_K$, one that must be
included now that results for $B_K$ have errors below the 10\% level~\cite{buras-2008}.
For completeness we note that the contributions of operators of
dimension-8 have been estimated to be very small, enhancing $B_K$ 
by $\sim 0.5\%$~\cite{cata-peris-2004}.

An accurate determination of $B_K$ thus constrains the CKM matrix,
as discussed here in the review of Van de Water~\cite{ruth-2009-1}.
%
Several groups are pursuing this calculation, using a variety of lattice
fermions~\cite{lubicz-2009-1}.
Since chiral symmetry constrains the four-fermion matrix element,
lattice fermions with some measure of chiral symmetry are preferred.
Our approach is to use improved staggered fermions, which are computationally
cheap, although one pays some price due to taste-breaking effects.
More precisely we use HYP-smeared valence staggered fermions (since
HYP-smearing is effective at reducing taste-breaking) and asqtad
staggered sea quarks (i.e. the MILC ensembles).
We take advantage of the exact $U(1)_A$ chiral symmetry and control
the systematic errors due to the taste symmetry breaking using 
staggered chiral perturbation theory
(SChPT)~\cite{ref:wlee:1,ref:bernard:1,ref:sharpe:1}.

This paper is the first report in a series of the four 
describing different aspects of our work.
Here we explain briefly the fitting procedures and results from
our SU(3) SChPT analysis.

\section{SU(3) Staggered ChPT Analysis}

In setting up a calculation of $B_K$ using
staggered fermions, one must deal with the
additional taste degree of freedom.
For the sea-quarks, this is done by using a rooted determinant.
For the valence quarks, one must choose a particular taste for the external
kaons, and choose the lattice operator so as to match the matrix
element onto the desired continuum one. Away from the continuum limit,
there are then several sources of error: taste-symmetry breaking coupled
with rooting leads to unphysical (unitarity violating) effects;
taste-symmetry breaking in the valence sector leads to significant changes
to the size of chiral logarithms; and approximate matching of the lattice
and continuum matrix elements (in our case using one-loop perturbation
theory) leads to truncation errors proportional to powers of $\alpha$.
All these effects are incorporated into SChPT,
which thus provides a method to systematically control and remove
these sources of error.
It  has been applied successfully to the analysis of the $\pi$, $K$, $D$
and $B$ meson masses
and decay constants, as reviewed in Ref.~\cite{ref:milc:2009}.

The theoretical SChPT analysis was generalized
to $B_K$ in Ref.~\cite{ref:sharpe:1}, and the result
worked out to next-to-leading order (NLO). This work assumed the
same staggered action for both valence and sea quarks and thus is
not directly applicable to our mixed-action setup.
We have carried out the appropriate generalization,
as summarized in Ref.~\cite{ref:wlee:2008-1}.
Full details will be presented in Ref.~\cite{ref:future}.

The power-counting used in the SChPT calculation is
$m_q/\Lambda_{\rm QCD}\sim (a\Lambda_{\rm QCD})^2
\sim \alpha/\pi \sim \alpha^2 $. Discretization errors
proportional to $a^2$ and to $a^2\alpha^2$ are treated
as of the same size, since, in practice, the latter (which
include taste-breaking effects) are numerically enhanced.
Since the dominant one-loop matching contributions are included,
and the coefficients of the taste-breaking operators known to be
small, these residual one-loop effects (denoted $\alpha/\pi$ above)
are treated as comparable to the unknown two-loop effects ($\alpha^2$ above).
For more discussion see Ref.~\cite{ref:sharpe:1}.

In the SU(3) SChPT analysis $m_q$ can be the mass of the light
or strange sea-quarks ($m_\ell$ and $m_s$, respectively), or the
mass of the down or strange valence quark ($m_x$ and $m_y$, respectively).
An important issue is whether $m_s/\Lambda_{\rm QCD}$ and
$m_y/\Lambda_{\rm QCD}$ are small enough for NLO ChPT to be useful. 
We address this phenomenologically
by including an analytic NNLO term.

At fixed $a$, the NLO expression contains 15 different low-energy
coefficients (LECs).\footnote{%
Not including $f_\pi$, which we fix at its experimental value,
taking the scale from the MILC collaboration.}
4 of these are present in the continuum, while the remaining 11
arise from taste-violating discretization and truncation errors
(including 2 which are introduced by the use of a mixed action).
Although we calculate $B_K$ for 55 different combinations of $m_x$
and $m_y$, and in some cases for several values of $m_\ell$,
it is not possible to do an unconstrained fit to the full functional form.
This is partly because the functions fall into groups within which
they are very similar (differing, for example, 
by the taste of the pion in the loops).
The differences within groups are comparable to or 
smaller than our statistical errors.
We have chosen to proceed by picking one of functions from each
group as a representative.
In addition, as described in Ref.~\cite{ref:wlee:2008-1},
the success of HYP smearing at reducing taste breaking implies
that two of the LECs are very small, and so we have dropped these terms.

After these choices, the number of parameters associated with
taste-breaking is reduced from 11 to 3, so the total number is reduced to 
the more manageable 7. We write the fitting function as:

\begin{eqnarray}
f_\text{th} &=& \sum_{i=1}^{7} c_i F_i
\label{eq:fth}
\end{eqnarray}
where $c_i$ are the coefficients (related to SChPT LECs), while 
the functions are
\begin{eqnarray}
  F_1 &=& 1 + \frac{1}{8\pi^2 f_\pi^2 G}
  \Big[ M_{\textrm{conn}} + M_{\textrm{disc}} \Big] \,,
  \qquad 
  F_2 = G / \Lambda^2 \,,
  \qquad
  F_3 = ( G / \Lambda^2 )^2 \,,
  \qquad
  F_4 = F^{(4)}_{\TT}
  \\
  F_5 &=& \frac{(X_\TP - Y_\TP)^2}{ G \Lambda^2 } \,,
  \qquad
  F_6 = F^{(6)}_{\TV\TA} \,,
  \qquad
  F_7 = F^{(1)}_\TA\,,
\end{eqnarray}
and depend on the various pseudo-Goldstone boson (PGB) masses.
Our notation is that $G\equiv K_\TP$ is 
the squared mass of the taste-$\xi_5$
valence PGB with flavor composition $xy$, $X_\TB$ and $Y_\TB$
are the squared masses of the taste-B valence PGBs with compositions 
$xx$ and $yy$, respectively, and $\Lambda_\chi\approx 4\pi f_\pi$
is the ChPT scale. We set $\Lambda_\chi=1\,$GeV in our fits---changing
this value results in shifts in the LECs.

The function $F_1$ contains the continuum-like chiral logarithms,
with discretization errors entering through the taste-breaking
in the masses of the PGBs in the loops. The quark-line connected part is:
\begin{equation}
  M_\text{conn} = \sum_{\TB=I,P,V,A,T} \frac{\tau^\TB}{2} 
\Big[ (G+X_\TB) \ell(X_\TB)
    + (G +Y_\TB)\ell(Y_\TB) + 2 (G-K_\TB) \ell(K_\TB)
    - 2 G K_\TB \tilde{\ell}(K_\TB) \Big]
\end{equation}
where $\tau^\TB$ is the fractional multiplicity of taste B,
and the chiral logarithmic functions are
\begin{eqnarray}
\ell(X) &=& X \log(X/\mu_{\rm DR}^2) \,,
\qquad
\tilde\ell(X) = - \frac{d\ell(X)}{dX} = -\log(X/\mu_{\rm DR}^2) -1\,,
\label{eq:logs}
\end{eqnarray}
up to (known) finite-volume corrections.
The expression for the quark-line disconnected part $M_\text{disc}$
is more lengthy: it can be deduced from the results 
in Ref.~\cite{ref:sharpe:1} and  
will be given in Ref.~\cite{ref:future}.
The important point is that, like $M_{\rm conn}$ 
it is fully predicted in terms of $f_\pi$ and
PGB masses that can be calculated in our simulation or obtained
from the results of the MILC collaboration.

The analytic terms $F_2$ and $F_5$ are also present in the
continuum, as is the NNLO analytic term $F_3$. Note that
$F_5$ vanishes when $m_x=m_y$.

The remaining three terms are due to taste-breaking
discretization or truncation errors, and are thus pure lattice artefacts.
Of the three, only $F_4$ is non-vanishing when $m_x=m_y$.
For brevity, we show only the two simplest functions:
\begin{eqnarray}
  F^{(4)}_{\TT} &=& \frac{3}{8} \frac{1}{f_\pi^2 G}
  \Big\{ 2 G \tilde{\ell}(K_\TT) +
  [\ell(X_\TT) + \ell(Y_\TT) - 2\ell(K_\TT)] \Big\}\,,
\\
  F^{(6)}_{\TV\TA} &=& \frac{3}{8} \frac{1}{f_\pi^2 G}
  \Big\{ [\ell(X_T)+\ell(Y_T) - 2 \ell(K_T)] \Big\}\,,
\end{eqnarray}
$F^{(1)}_{\TA}$ can be deduced from Ref.~\cite{ref:sharpe:1} and
will be given in Ref.~\cite{ref:future}.

\section{Fitting procedure}
%
%
%
\begin{table}[t!]
\centering
\begin{tabular}{c  c  c  c  l }
\hline
\hline
$a$ (fm) & $am_l/am_s$ & geometry & ens $\times$ meas & ID \\
\hline
0.12 & 0.03/0.05  & $20^3 \times 64$ & $564 \times 1$ & C1 \\
0.12 & 0.02/0.05  & $20^3 \times 64$ & $486 \times 1$ & C2 \\
0.12 & 0.01/0.05  & $20^3 \times 64$ & $671 \times 9$ & C3 \\
0.12 & 0.01/0.05  & $28^3 \times 64$ & $274 \times 8$ & C3-2 \\
0.12 & 0.007/0.05 & $20^3 \times 64$ & $651 \times 1$ & C4 \\
0.12 & 0.005/0.05 & $24^3 \times 64$ & $509 \times 1$ & C5 \\
\hline
0.09 & 0.0062/0.031 & $28^3 \times 96$ & $995 \times 1$ & F1 \\
\hline
0.06 & 0.0036/0.018 & $48^3 \times 144$ & $513 \times 1$ & S1 \\
\hline
\hline
\end{tabular}
\caption{MILC lattices used for the numerical study.
  Here ``ens'' and ``meas'' are the number of gauge configurations
  and measurements per configuration, respectively, and ``ID'' is
  a label.
  \label{tab:milc-lat}}
\end{table}
We calculate $B_K$ using standard methods~\cite{msBKpaper}
on the MILC lattices listed in Table~\ref{tab:milc-lat}.
We use the 10 valence quark masses given in Table \ref{tab:val-qmass},
and thus have 55 different mass combinations for the PGBs:
10 degenerate and 45 non-degenerate.
We use one-loop matching.

\begin{table}[htbp]
\centering
\begin{tabular}{ c  c  c}
\hline
\hline
$a$ (fm) & $a m_x$ and $a m_y$ & $n$ \\
\hline
0.12 &  $0.005 \times n$  & $1,2,3,\ldots,10$ \\
0.09 &  $0.003 \times n$  & $1,2,3,\ldots,10$ \\
0.06 &  $0.0018 \times n$ & $1,2,3,\ldots,10$ \\
\hline
\hline
\end{tabular}
\caption{Valence quark masses used for the numerical study
(with $m_x\le m_y$).
  \label{tab:val-qmass}}
\end{table}
%

%
\begin{figure}[t!]
\centering
    \includegraphics[width=0.45\textwidth]
                    {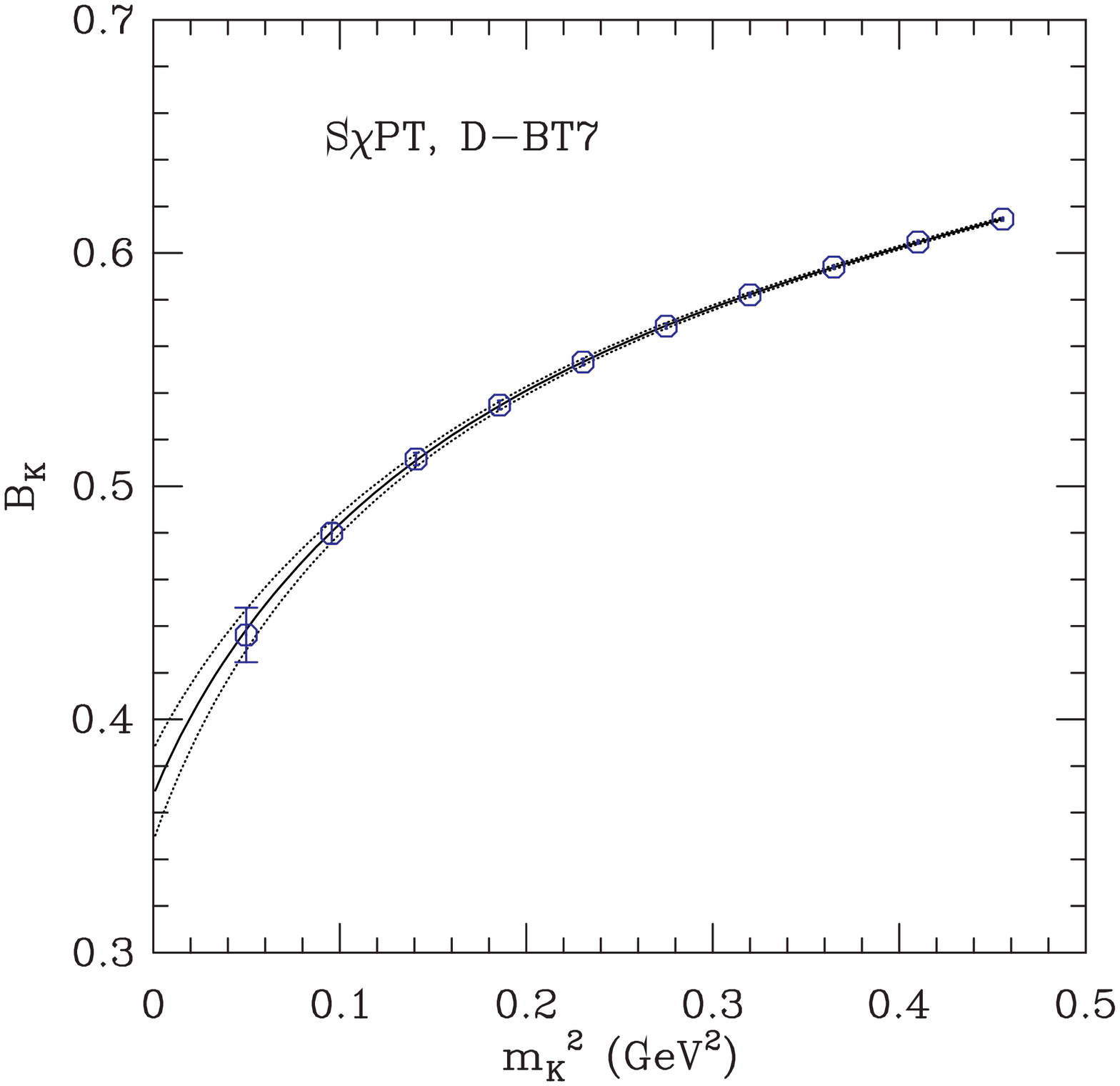}
    \includegraphics[width=0.49\textwidth]
                    {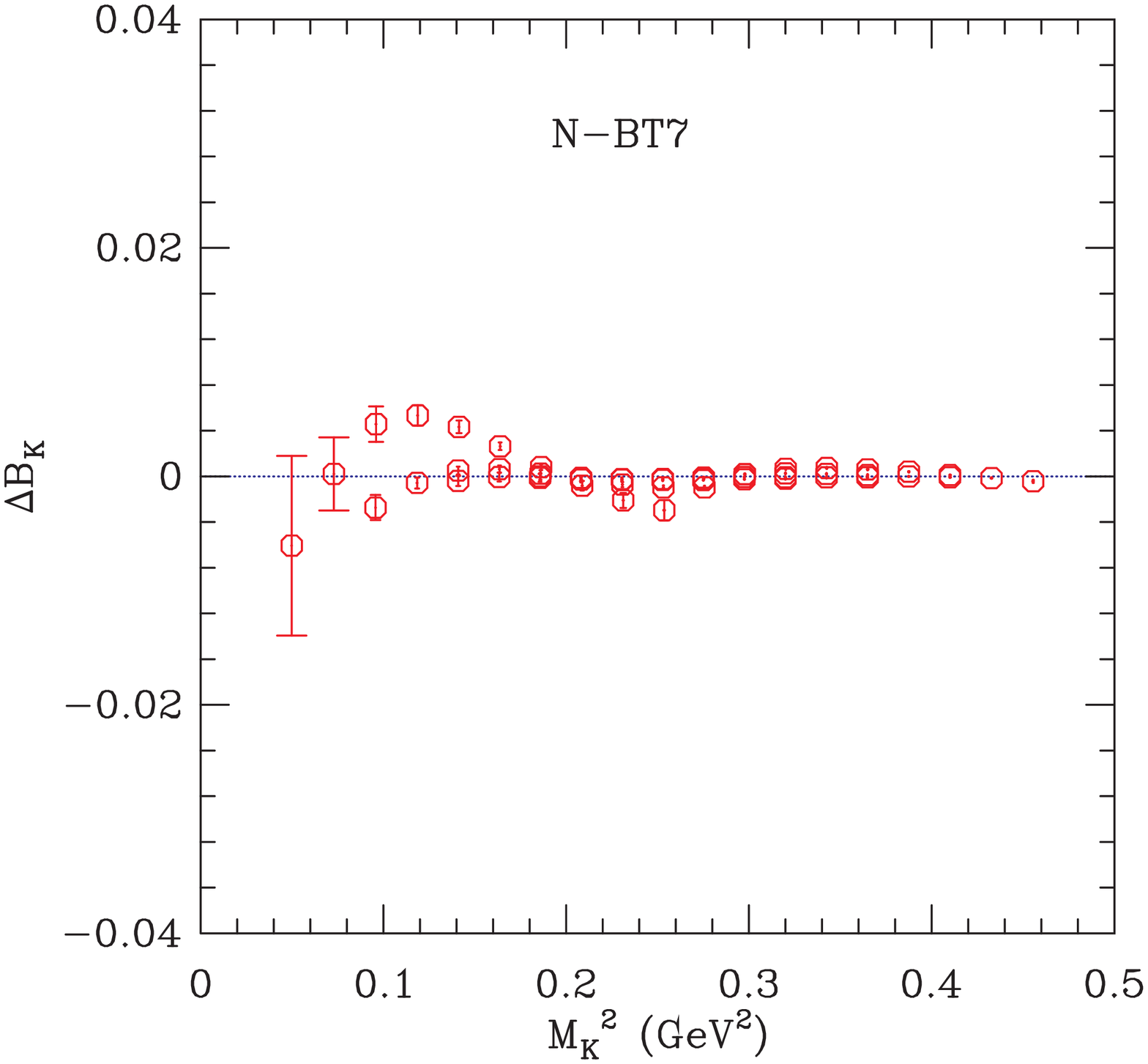}
\caption{ $B_K$ vs. $m_K^2$ (left) and $\Delta B_K$ vs. $m_K^2$
  (right): The results are on the C3 lattices. The fit
  types are D-BT7 (left, degenerate case) and N-BT7 (right,
  non-degenerate case). 
}
\label{fig:sxpt:bt7}
\end{figure}
We use several fitting procedures, but focus here on the one
used for our central value. In this fit we do not yet include
the finite-volume corrections predicted by ChPT.
First, we fit the degenerate data. The general form (\ref{eq:fth}) 
reduces to $\sum_{i=1}^{4} c_i F_i$ in the degenerate limit.
We know  the expected size of each term; in particular, we
expect $c_4$ to have the magnitude $c_4 \approx a^2 \Lambda_\textrm{QCD}^4$
if it is dominated by discretization errors.
We implement this prior knowledge using Bayesian fitting~\cite{ref:bayesian}
in which the standard $\chi^2$ is augmented by the addition of 
\begin{equation}
\chi^2_\text{prior} = \frac{( c_4 - a_4)^2}{\tilde{\sigma}^2_4}\,.
\end{equation}
We take $a_4=0$ since we do not know the sign of $c_4$,
and $\tilde{\sigma}_4=a^2 \Lambda_\textrm{QCD}^4$ with
$\Lambda_\textrm{QCD}=0.3\,$GeV.
Minimizing the augmented $\chi^2$ leads to the fit we label
D-BT7 (with ``D'' for degenerate).\footnote{%
At the present stage, our $\chi^2$ includes only
the diagonal elements of the correlation matrix.}
An example is shown in the left panel of Fig.~\ref{fig:sxpt:bt7}.

Next we fit $f_{\rm th}$ to all 55 data points, but using
the results for $c_{1-4}$ and their errors from the D-BT7 fit 
as priors for the first four parameters.
We also include priors for the coefficients $c_6$ and $c_7$,
whose expected magnitudes are
$c_6 \approx c_4$ and $c_7 \approx \Lambda_{\rm QCD}^2 c_4$.
In total then, $\chi^2$ is augmented by
\begin{equation}
\sum_{i=1}^{4} \frac{( c_i - a_i)^2}{\tilde{\sigma}^2_i}
+
\sum_{i=6,7} \frac{( c_i - a_i)^2}{\tilde{\sigma}^2_i}
\end{equation}
with $a_{1-4}$ set to the central values of $c_{1-4}$ from the D-BT7 fit,
and $\tilde{\sigma}_{1-4}$ set to the corresponding errors, while
$a_6=a_7=0$, $\tilde{\sigma}_6=a^2 \Lambda_\textrm{QCD}^4$
and $\tilde{\sigma}_7=a^2 \Lambda_\textrm{QCD}^6$.
The resulting fit is labeled N-BT7 (``N'' for non-degenerate).
An  example is shown in the right panel of Fig.~\ref{fig:sxpt:bt7},
where $\Delta B_K$ is the residual:
\begin{equation}
\Delta B_K \equiv B_K(\text{data}) - f_\text{th}\,.
\end{equation}
The fit works well except for the non-degenerate points in which
$m_x$ takes its smallest value, which we suspect is due in part
due to finite-volume effects.

We have tried many variations of this fitting strategy:
relaxing the constraints on $c_{4,6,7}$ to the values one might
expect if $O(\alpha^2)$ [rather than $O(a^2)$] 
effects were dominant (type N-BT7-2);
fitting in one stage to all data-points with either
$O(a^2)$ or $O(\alpha^2)$ priors on $c_{4,6,7}$ (types N-BT8 and N-BT8-2);
and simply fitting without priors (N-T2).
The results for the continuum-like parameters $c_{1-3}$ and $c_5$
are consistent between fits,
%
%
while those for the lattice artefacts $c_4$, $c_{6-7}$ vary
significantly.  An intriguing result is that we find $c_5$ to be
remarkably smaller than the expected size [$c_5\sim O(1)$].

For each fit, we determine a value for $B_K$ at the
physical valence-quark masses by setting all lattice
artefacts to zero in the SChPT expression. We can then
attempt to extrapolate to the continuum 
and infinite volume limits, and to the physical
values of $m_\ell$ and $m_s$. Examples of some of
these extrapolations are shown in the companion 
proceedings~\cite{ref:wlee:2009-2,ref:wlee:2009-3}.

\section{Error budget and conclusion}
\begin{table}[htbp]
\centering
\begin{tabular}{ l | l l }
\hline \hline
source & error (\%) & description \\
\hline
statistics       & 2.0   & N-BT7 fit \\
discretization   & 2.5   & diff.~of (S1) and ($a=0$)\\
fitting (1)      & 0.6   & diff.~of N-BT7 and N-BT8 (C3) \\
fitting (2)      & 5.4   & diff.~of N-BT7 and N-BT7-2 (C3) \\
fitting (3)      & 1.3   & fit w.r.t. $am_l$ \\
finite volume    & 2.2   & diff.~of C3 and C3-2 \\
matching factor  & 6.3   & $\Delta B_K^{(2)'}$ (S1) \\
scale $r_1$      & 0.07  & uncertainty in $r_1$ \\
\hline \hline
\end{tabular}
\caption{Preliminary error budget for $B_K$ obtained using SU(3) SChPT fitting.
  \label{tab:su3-err-budget}}
\end{table}
In Table \ref{tab:su3-err-budget}, we summarize our present best
estimates of the uncertainties in $B_K$ coming from various sources.
These are conservative estimates, some of which may decrease with
further analysis of our present data set. 
The method by which we estimate these errors
is outlined in the ``description''.
The first fitting error estimates the (rather small)
impact of fitting in two stages rather than in one, 
while the second fitting error estimates the impact of changing the widths
$\tilde\sigma_{4,6,7}$ for the priors. The latter error is one of our two
dominant systematic errors---it indicates that we cannot pin down
the functional form of $B_K$ well enough to extrapolate to
the physical light-quark mass with less than 5\% uncertainty
(at least on the coarse lattices).
The other dominant error is that due to the truncation of
the matching factor at one-loop order. This error is discussed
in one of the companion proceedings~\cite{ref:wlee:2009-4}.

Our current, \textbf{preliminary}
estimate of $B_K$ using SU(3) ChPT fitting is
\begin{equation}
B_K(\text{NDR},\mu=2\text{ GeV}) = 0.528 \pm 0.011 \pm 0.048
\qquad \textrm{[SU(3), \textbf{PRELIMINARY}.]}
\end{equation}
where the first error is statistical and the second combines
all systematic errors.
This turns out to be less accurate than the 
results from SU(2) ChPT fitting~\cite{ref:wlee:2009-2}.

\section{Acknowledgments}
C.~Jung is supported by the US DOE under contract DE-AC02-98CH10886.
The research of W.~Lee is supported by the Creative Research
Initiatives Program (3348-20090015) of the NRF grant funded by the
Korean government (MEST). 
The work of S.~Sharpe is supported in part by the US DOE grant
no.~DE-FG02-96ER40956. Computations were carried out
in part on facilities of the USQCD Collaboration,
which are funded by the Office of Science of the
U.S. Department of Energy.

\end{document}